\documentclass[1p]{elsarticle}

\usepackage{hyperref}
\usepackage{graphicx}
\usepackage{amsfonts}
\usepackage{amsmath}
\usepackage{amssymb}
\usepackage{amsthm} 
\usepackage[inline]{enumitem}
\usepackage{algorithm}
\usepackage[noend]{algpseudocode} 

\usepackage[colorinlistoftodos]{todonotes}
\newtheoremstyle{def}
{3pt}       
{3pt}       
{}          
{}          
{\scshape}  
{.}         
{.5em}      
{}          

\theoremstyle{def}
\newtheorem{definition}{Definition}
\newtheorem{theorem}{Theorem}
\newtheorem{lemma}{Lemma}
\newtheorem{problem}{Problem}

\journal{Journal of Pattern Recognition}

\begin{document}

\begin{frontmatter}

\title{Separating Structure from Noise in Large Graphs \\ Using the Regularity Lemma}


\author[DAIS]{Marco Fiorucci\corref{mycorrespondingauthor}}
\cortext[mycorrespondingauthor]{Corresponding author}
\ead{marco.fiorucci@unive.it}
\author[DAIS]{Francesco Pelosin}
\author[DAIS,ECLT]{Marcello Pelillo}

\address[DAIS]{DAIS, Ca’ Foscari University, Via Torino 155, Venice, Italy}
\address[ECLT]{ECLT, Ca’ Foscari University, S. Marco 2940, Venice, Italy}

\begin{abstract}
How can we separate structural information from noise in large graphs? To address this fundamental question, we propose a graph summarization approach based on Szemer\'edi's Regularity Lemma, a well-known result in graph theory, which roughly states that every graph can be approximated by the union of a small number of random-like bipartite graphs called ``regular pairs''. Hence, the Regularity Lemma provides us with a principled way to describe the essential structure of large graphs using a small amount of data. Our paper has several contributions: (i) We present our summarization algorithm which is able to reveal the main structural patterns in large graphs. (ii) We discuss how to use our summarization framework to efficiently retrieve from a database the top-$k$ graphs that are most similar to a query graph. (iii) Finally, we evaluate the noise robustness of our approach in terms of the reconstruction error and the usefulness of the summaries in addressing the graph search task.
\end{abstract}


\end{frontmatter}

\section{Introduction}
Recent years are characterized by an unprecedented quantity of available network data which are produced at an astonishing rate by an heterogeneous variety of interconnected sensors and devices. This high-throughput generation calls for the development of new effective methods to store, retrieve, understand and process massive network data. 
To tackle this challenge, we introduce a framework for summarizing large graphs based on \emph{Szemer\'edi's Regularity Lemma}\cite{Szemeredi75Graphs}, which roughly states that any sufficiently large graph can almost entirely be partitioned into a bounded number of random-like bipartite graphs, called \emph{regular pairs}. The partition resulting from the Regularity Lemma gives rise to a summary, called \emph{reduced graph}, which inherits many of the essential structural properties of the original graph \cite{Komlos2002,Komlos1996}. 

In this paper, we posit that the Regularity Lemma can be used to summarize large graphs revealing its main structural patterns, while filtering out noise, which is common in any real-world networks.
In its original form, the lemma is an existential predicate, but during the last decades various constructive algorithms have been proposed \cite{Alon94,CzygrinowRödl,Frieze1999APartition}. However, despite being polynomial  in  the  size  of  the underlying  graph, all these algorithms have a  hidden tower-type dependence on an accuracy parameter. To overcome this limitation, in the last years we have proposed some simple heuristics that, most of the times, allowed us to construct a regular partition \cite{Fiorucci17,PelilloRevealing17,Sperotto07}.

The main contribution of this paper is the introduction of a new heuristic algorithm which is characterized by an improvement of the summary quality both in terms of reconstruction error and of noise filtering. In particular, we first build the reduced graph of a graph $G$, and then we "blow-up" the reduced graph to obtain a graph $G'$, called \emph{reconstructed graph}, which is close to $G$ in terms of the $l_p$-reconstruction error. We study the noise robustness of our approach in terms of the reconstruction error by performing an extensive series of experiments on both synthetic and real-world data. As far as the synthetic data are concerned, we generate graphs with a cluster structure, where the clusters are perturbed with different levels of noise. As far as the real-world data are concerned, we add spurious edges in accord with different noise probabilities. The aim of this series of experiments is to assess if the framework is able to separate structure from noise. In the ideal case, the distance between $G$ and $G'$ should be only due to the filtered noise.

Moreover, in the second part of the paper, we use our summarization algorithm to address the \emph{graph search} problem defined under a similarity measure. The aim of graph search is to retrieve from a database the top-$k$ graphs that are most similar to a query graph. Since noise is common in any real-world dataset, the biggest challenge in graph search is developing efficient algorithms suited for dealing with large graphs containing noise in terms of missing and adding spurious edges. 
In our approach, all the graphs contained in a database are compressed off-line, while the query graph is compressed on-line. Thus, graph search can be performed on the summaries, and this allows us to speed up the search process and to reduce storage space. Finally, we evaluate the usefulness of our summaries in addressing the graph search problem by performing an extensive series of experiments. In particular, we study the quality of the answers in terms of the found top-$k$ similar graphs, and the scalability both in the size of the database and in the size of the query graphs.


\paragraph{\textbf{Related Works}}
The first contribution of our paper is the introduction of a principled framework for summarizing large graphs with the aim of preserving their main structural patterns. Previous related works presented methods which mainly built summaries by grouping the vertices into subsets, such that the vertices within the same subset share some topological properties. The works in \cite{SCHAEFFER200727,Newman} introduced methods for partitioning the vertices into non-overlapping clusters, so that vertices within the same cluster are more connected than vertices belonging to different clusters. A graph summary can be constructed by considering each cluster as a \emph{supernode}, and by connecting each pair of supernodes with a \emph{superedge} of weight equals to the sum of the cross-cluster edges. However, since graph summarization and clustering have different goals, this approach is suited only if the input graph has a strong community structure.
In \cite{LeFreve}, the summary is generated by greedily grouping vertices, such that the normalized reconstruction error between the adjacency matrix of the input graph and the adjacency matrix of the \emph{reconstructed graph} is minimized. Since in their work they exploited heuristic algorithms, they can not give any guarantees on the quality of the summary. The work in \cite{Riondato2017} proposed a method of building a summary with quality guaranty by minimizing the \emph{$l_p$-reconstruction error} between the adjacency matrix of the input graph and the adjacency matrix of the reconstructed graph. Since both approaches aim to minimize a distance measure between the input and the reconstructed graph, they are not the best choice for summarizing noisy graphs. By contrast, our goal is to develop a graph summarization algorithm which is robust against noise.
For a more detailed picture on how the field has evolved previously, we refer the interested reader to the survey of Liu et al. \cite{Liu2017GraphSurvey}.

The second contribution of our paper consists in addressing the graph search problem using the proposed summarization framework. Locating the occurrences of a query graph in a large database is a problem which has been approached in two main different ways, based on subgraph isomorphism and approximate graph matching respectively.
Ullman \cite{Ullmann76} put one of the first milestones in subgraph isomorphism. He proposed an algorithm which decreases the computational complexity of the matching process by reducing the search space with backtracking. Recently, Carletti et al. \cite{Carletti17} introduced an algorithm for graph and subgraph isomorphism  which scales better than Ullmann's one. In particular, Carletti et al's algorithm, which may be considered as the state-of-the-art in exact subgraph matching, can process graphs of size up to ten thousand nodes. However, since subgraph isomorphism is a NP-complete problem, the algorithms based on exact matching are prohibitively expensive for querying against a database which contains large graphs. Moreover, due to the noise contained in any real-world networks, it is common to mismatch two graphs which have the same structure but different levels of noise. Indeed, these contributions are focused on exact matching and, even if they proposed efficient solutions, they are not noise robust. By contrast, our goal is to develop an efficient graph search algorithm which is robust against noise. 
Hence, approaches based on approximate graph matching are more suitable for addressing the graph search problem. Indeed, in this category lies the most effective graph similarity search algorithms. Most of the time, the searching phase is conducted under the \emph{graph edit distance} ($GED$) constraint \cite{Liang17,Zheng15,Zheng13}. The graph edit distance $GED(g_1,g_2)$ is defined as the minimum number of edit operations (adding, deletion and substitution) that modify $g_1$ step-by-step to $g_2$ (or vice versa). In \cite{Zheng15} and in \cite{Zhang10}, the authors underline the robustness of $GED$ against noise due to its error-tolerant capability. Unfortunately, the $GED$ computation is NP-hard, and most existing solutions adopt a \emph{filtering-verification} technique. In particular, first, a pruning strategy is used to filter out false positive matches, and then the remaining candidates are verified by computing $GED$. In this context, the work of Liang and Zhao \cite{Liang17} represents the state-of-the-art. They provided a partition-based $GED$ lower bound to improve the filter capability, and a multi-layered indexing approach to filter out false positives in an efficient way. Their algorithm can deal with databases with a high number of graphs, but cannot handle large graphs due to the complexity of $GED$ computation. Instead, our algorithm is designed to scale both in the size of the databases and in the size of the graphs.
\paragraph{\textbf{Roadmap}}
The paper is organized as follows. In section \ref{preliminaries}, we provide the basic concepts and notations used in the next sections as well as the formal definition of graph summary. In section \ref{RL}, we present a short theoretical description of Szemer\'edi's Regularity Lemma and we describe how to find a regular partition. Section \ref{sumAlg} is devoted to the description of our summarization framework, while in section \ref{graphSearch}, we introduce the formal definition of graph search problem, and we discuss how to use our framework to speed up the search process and to reduce the storage space. In section \ref{experiments}, we report an extensive experimental evaluation of the noise robustness of our approach, in terms of the reconstruction error, and of the usefulness of the summaries in addressing the graph search task. Finally, we drawn our conclusions in section \ref{conclusions}.

\section{Preliminaries}\label{preliminaries}
Let $G = (V,E)$ be an undirected graph without self-loops. The \emph{edge density} of a pair of two disjoint vertex sets 
$C_i, C_j \subseteq V$ is defined as: 
\begin{equation}
d(C_i,C_j) = \frac{e(C_i,C_j)}{\left\vert{C_i}\right\vert \left\vert{C_j}\right\vert }
\end{equation}
where $e(C_i, C_j)$ denotes the number of edges of $G$ with an endpoint in $C_i$ and an endpoint in $C_j$.

Given a positive constant $\varepsilon > 0$, we say that the pair $(C_i,C_j)$ of disjoint vertex sets $ C_i,C_j \subseteq V $ is {\em $\varepsilon$-regular}
if for every $ X \subseteq C_i $ and $ Y \subseteq C_j $ satisfying 
$\left\vert{X}\right\vert > \varepsilon \left\vert{C_i}\right\vert \mbox{and} \left\vert{Y}\right\vert > \varepsilon \left\vert{C_j}\right\vert$
we have
\begin{equation}
\left\vert{d(X,Y) - d(C_i,C_j)}\right\vert < \varepsilon~.
\end{equation}
This means that the edges in an $\varepsilon$-regular pair are distributed fairly uniformly, where the deviation from the uniform distribution is controlled by the tolerance parameter $\varepsilon$.

\begin{figure}[t!]
\centering
\includegraphics[width=0.8\textwidth]{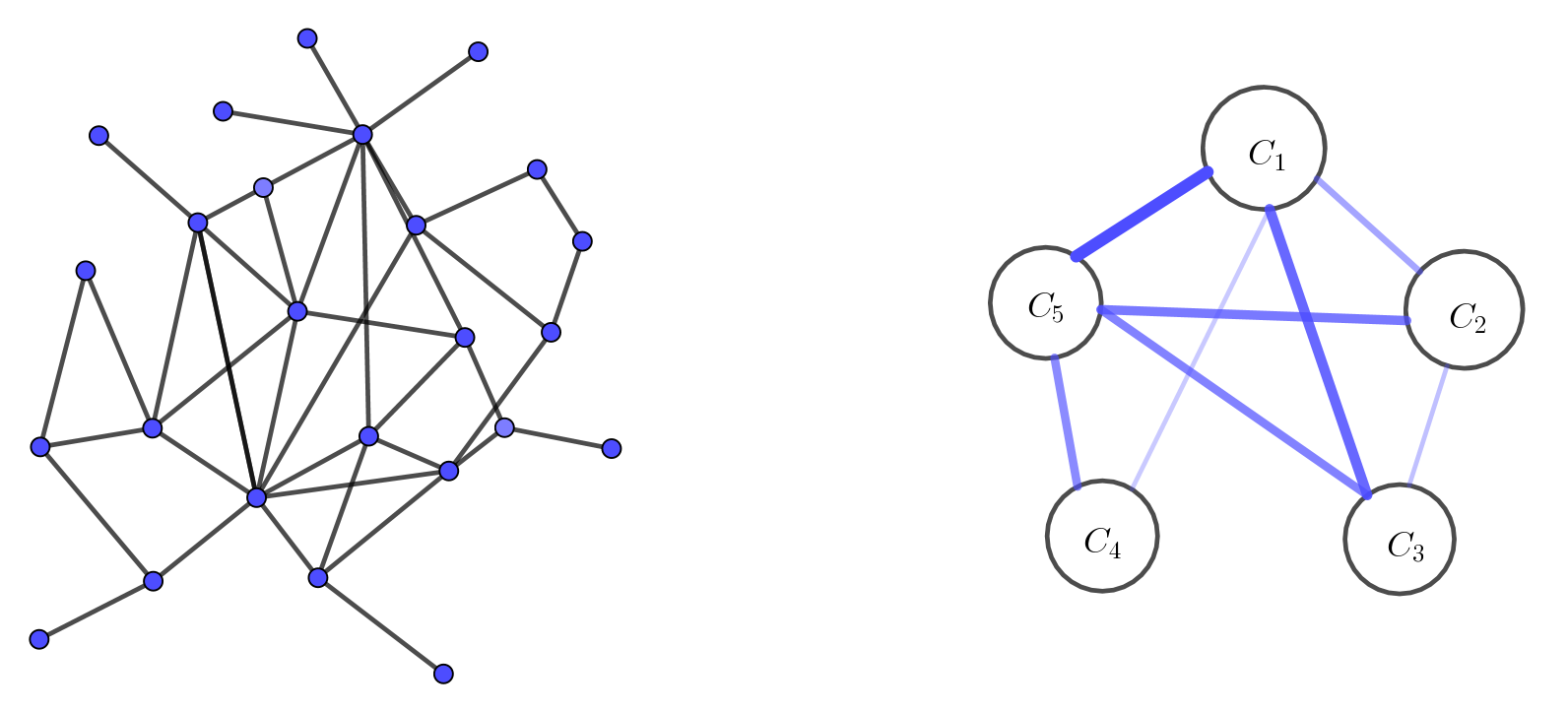}
\caption{\label{fig:approximation}Example of the reduced graph (summary) construction. Left: the original graph. Right: the \emph{reduced graph} which contains eight $\varepsilon$-regular classes pairs. The density of each pair is expressed by the thickness of the edge that connects the classes of that pair. If a pair is $\varepsilon$-irregular the corresponding classes are not connected by an edge.}
\end{figure}

\begin{definition}[$\varepsilon$-regular partition]
A partition $\mathcal{P} = \{C_0, C_1, \dots , C_k\}$, with $C_0$  being the exceptional set\footnote{$C_0$ has only a technical purpose: it makes it possible that all other classes have exactly the same number of vertices.} is called \emph{$\varepsilon$-regular} if:
\begin{enumerate}
\item it is equitable: $|C_1| = |C_2| = \dots = |C_k|$; 
\item $|C_0| < \varepsilon|V|$;
\item all but at most $\varepsilon k^2$ of the pairs $(C_i, C_j)$  are $\varepsilon$-regular, ($1 \leq i < j \leq k$).
\end{enumerate}
\end{definition}

In this paper, we propose a summarization algorithm which, given an undirected graph $G=(V,E)$, iteratively builds a summary, called \emph{reduced graph} (Figure \ref{fig:approximation}), defined as follows.
\begin{definition}[Reduced graph]	
Given an $\varepsilon$-regular partition $P = \{ C_{1}, C_{2},$\\$\dots, C_{k} \}$ of a graph $G = (V, E)$ and $0 \leq d' \leq 1$, the reduced graph of $G$ is the undirected weighted graph $R = (V_{R}, E_{R}, w)$, where $V_{R} = P$, $E_{R} \subseteq V_{R}^{2}$ and $w: E_{R} \rightarrow \mathbb{R}$ is defined as follows:

\[ w((C_{i}, C_{j})) = 
\begin{cases}
d(C_{i}, C_{j}) & \text{if } (C_{i}, C_{j}) \text{ is $\varepsilon$-regular and } d(C_{i}, C_{j}) \geq d', \\
0 & \text{otherwise.}
\end{cases}
\]
\end{definition}

We are now ready to state the Regularity Lemma which provides us a principled way to develop a summarization algorithm with the aim of separating structure from noise in a large graph. 
\section{Szemer\'edi's Regularity Lemma} \label{RL}
In essence, Szemer\'edi's Regularity Lemma states that given an $\varepsilon > 0$, every sufficiently dense graph $G$ can be approximated by the union of a bounded number of quasi-random bipartite graphs, where the deviation from randomness is controlled by the tolerance parameter $\varepsilon$. In other words, we can partition the vertex set $V$ into a bounded number of classes $C_0, C_1, . . . , C_k$, such that almost every pair $(C_i, C_j)$ behaves similarly to a random bipartite graph, ($1 \leq i < j \leq k$).
\begin{lemma}[Szemer\'edi's regularity lemma ($1976$)]
  For every positive real $\varepsilon$ and for every positive integer $m$, there are positive integers $N = N(\varepsilon,m)$ and $M = M(\varepsilon,m)$ with the following property: for every graph $G=(V,E)$, with $\left\vert{V}\right\vert \geq N$, there is an $\varepsilon$-regular partition of $G$ into $k + 1$ classes such that $m \leq k \leq M$.
\end{lemma}

The strength of the Regularity Lemma is corroborated by the so-called Key Lemma, which is an important theoretical result introduced by Komlos et al. \cite{Komlos1996}. It basically states that the reduced graph does inherit many of the essential structural properties of the original graph. 
Before presenting its original formulation, another kind of graph needs to be defined, namely the \emph{fold graph}.
Given an integer $t$ and a graph $R$ (which may be seen as a reduced graph), let $R(t)$ denote the graph obtained by ``blowing up'' each vertex $j$ of $V(R)$ to a set $A_j$ of $t$ independent vertices, and joining $u \in A_x$ to $v \in A_y$ if and only if $(x,y)$ is an edge in $R$.
Thus, $R(t)$ is a graph in which every edge of $R$ is replaced by a copy of the complete bipartite graph $K_{tt}$.
The following lemma shows a link between the reduced graph $R$ and $R(t)$. 

\begin{theorem}[Key Lemma]
\label{KeyLemma}
Given $d > \varepsilon > 0$, a graph $R$, and a positive integer $m$, let us construct a graph $G$ by performing the following steps:
\begin{enumerate}
\item replace every vertex of $R$ by $m$ vertices;
\item replace the edges of $R$ with $\epsilon$-regular pairs of density at least $d$.
\end{enumerate}
Let $H$ be a subgraph of $R(t)$ with $h$ vertices and maximum degree $\Delta >0$, and let $\delta = d-\varepsilon$ and $\varepsilon_0 = \delta^\Delta/(2+\Delta)$. If $\varepsilon \leq \varepsilon_0$ and $t-1 \leq \varepsilon_0m$, then $H$ is embeddable into $G$ (i.e. $G$ contains a subgraph isomorphic to $H$).
In fact, we have:
\begin{equation}
\left\vert\left\vert{H\,\to\,G}\right\vert\right\vert > (\varepsilon_0m)^h
\end{equation}
where $\left\vert\left\vert{H\,\to\,G}\right\vert\right\vert$ denotes the number of labeled copies of $H$ in $G$.
\end{theorem}

Hence, the Key Lemma provides us a theoretical guarantee on the quality of the summary built from an $\epsilon$-regular partition. In particular, for $t=1$, $R(t)=R$, and if the constraint on the edge density $d$ is satisfied, the Key Lemma ensures that every small subgraph of $R$ is also a subgraph of $G$. Thus, we can use the Regularity Lemma to build a summary $R$ of $G$, and then we can infer structural properties of $G$ by studying the same properties on $R$.  

\paragraph{\textbf{Finding Regular Partitions}}
The original proof of the Regularity Lemma is not constructive, but during the last decades different constructive versions have been proposed. In this paper, we focus on the Alon et al.'s \cite{Alon94} work. In particular, they proposed a new formulation of the Regularity Lemma which emphasizes the algorithmic nature of the result.

\begin{theorem}{(Alon et al., 1994)}
For every $\varepsilon > 0$ and every positive integer $t$ there is an integer $Q = Q(\varepsilon, t)$ such that every graph with $n > Q$ vertices has an $\varepsilon$-regular partition into $k + 1$ classes, where $t \leq k \leq Q$. For every fixed $\varepsilon > 0$ and $t \geq 1$ such partition can be found in $O(M(n))$ sequential time, where $M(n) = O(n^{2.376})$ is the time for multiplying two $n \times n$ matrices with $0,1$ entries over the integers. It can also be found in time $O(\log{n})$ on an Exclusive Read Exclusive Write Parallel Random Access Machine(EREW PRAM) with a polynomial number of parallel processors.
\end{theorem}

A sketch of the proof is then presented. Let $H$ be a bipartite graph with color classes $A$ and $B$, with  $\left\vert{A}\right\vert = \left\vert{B}\right\vert = n$. Let us define the average degree $\bar{d}$ of $H$ as:
$$\bar{d}(A,B)=\frac{1}{2n} \sum_{i \in A \cup B} deg(i)$$
where $deg(i)$ is the degree of vertex $i$.

For two distinct vertices $y_1, y_2 \in B$ the \emph{neighbourhood deviation} of $y_1$ and $y_2$ is defined as:
$$\sigma(y_1, y_2) = |N(y_1) \cap N(y_2)| - \frac{\bar{d}^2}{n}$$

where $N(x)$ is the set of neighbours of vertex $x$. For a subset $Y \subset B$ the \emph{deviation} of $Y$ is defined as:
$$\sigma(Y)=\frac{\sum_{y_1,y_2 \in Y}\sigma(y_1,y_2)}{|Y|^2}$$\label{sigma}

Let $0 < \varepsilon < 1/16$, it can be proved that, if there exists $Y \subset B,\; |Y| > \varepsilon n$ such that $\sigma(Y) \geq  \frac{\varepsilon^3}{2} n$, then at least one of the following cases occurs:

\begin{enumerate}
\item $\bar{d} <\varepsilon^3 n$ ($H$ is $\varepsilon$-regular); \label{cond1}
\item there exists in $B$ a set of more than $\frac{1}{8}\varepsilon^4n$ vertices whose degrees deviate from $\bar{d}$ by at least $\varepsilon^4n$ ($H$ is $\varepsilon$-irregular); \label{cond2}
\item there are subsets $A'\subset A, \; B'\subset B, \; |A'|\geq \frac{\varepsilon^4}{n}n, \; |B'|\geq \frac{\varepsilon^4}{n}n$ such that $|\bar{d}(A', B') - \bar{d}(A,B)|\geq \varepsilon^4$ ($H$ is $\varepsilon$-irregular). \label{cond3}

\label{alon3}
\end{enumerate}

Note that one can easily check if \ref{cond1} holds in time $O(n^2)$. Similarly, it is trivial to check if \ref{cond2} holds in $O(n^2)$ time, and in case it holds to exhibit the required subset of $B$ establishing this fact.
If the first two conditions are not verified, the \ref{cond3} condition must be checked. To this end, we have to find the subsets $A',B'$, called \emph{certificates}, that witness the irregularity of the bipartite graph $H$. To address this task, we first select a subset of $B$ whose vertex degrees ``deviate'' the most from the average degree $\bar{d}$ of $H$. More formally: for each $y_0 \in B$ with $|deg(y_0)-\bar{d}| < \varepsilon^4n$ we find the vertex set $B_{y_0} = \{y \in B| \, \sigma(y_0, y) \geq 2\varepsilon^4 n \}$. The proof provided by Alon et al. guarantees the existence of at least one such $y_0$ for which $|B_{y_0}| \geq \frac{\varepsilon^4}{4}n$. Thus, the subsets $B'=B_{y_0}$ and $A'=N(y_0)$ are the required certificates. These two subsets represent the collection of vertices that contribute more to the irregularity of the pair $(A,B)$. The sets $\bar{A^{'}} = A \setminus A^{'}, \, \bar{B^{'}} = B \setminus B^{'}$ are called \emph{complements}. Since the computation of the quantities $\sigma(y, y')$, for $y, \; y' \in B$, can be done by squaring the adjacency matrix of $H$, the overall complexity of this algorithms is $O(M(n)) = O(n^{2.376})$.

Before reporting Alon et al.'s algorithm, we present a measure to assess the goodness of a partition of the vertex set $V$ of a graph $G$, which was introduced by Szemer\'edi \cite{Szemeredi75Graphs}. Given a partition $\mathcal{P} = \{C_0,C_1,\dots, C_k\}$ of a graph $G=(V,E)$, the index ($\emph{sze\_idx}$) of the partition $\mathcal{P}$ is defined as follows:

$$sze\_ind(\mathcal{P})=\frac{1}{k^2}\sum_{s=1}^{k}\sum_{t=s+1}^{k}d(C_s,C_t)^2$$\label{indexP}

Since $0\leq d(C_s,C_t) \leq 1$, it can be seen that $sze\_ind(\mathcal{P}) \leq \frac{1}{2}$.
Szemer\'edi proved that if a partition $\mathcal{P}$ violates the regularity condition, then it can be refined by a new partition $\mathcal{P '}$ such that $sze\_ind(\mathcal{P '}) > sze\_ind(\mathcal{P})$. Finally, we can now present Alon et al.'s algorithm, which provides a way to find an $\varepsilon$-regular partition. The procedure is divided into two main steps: in the first step all the constants needed during the next computation are set; in the second one, the partition is iteratively created. An iteration is called \textit{refinement step}, because, at each iteration, the current partition is closer to a regular one.

\paragraph{Alon et al.'s Algorithm}
\begin{enumerate}
\item Create the initial partition: arbitrarily divide the vertices of $G$ into an equitable partition $\mathcal{P}_1$ with classes $C_0,C_1, \dots, C_b$ where $|C_i|=\lfloor \frac{n}{b}\rfloor$. \label{step1}
\item Check Regularity: for every pair $(C_r,C_s)$ of $\mathcal{P}_i$, verify if it is $\varepsilon$-regular or find two certificates $A^{'} \subset C_r, \, B^{'} \subset C_s, \, |A^{'}| \geq \frac{\varepsilon^4}{16}|C_1|, \, |B^{'}| \geq \frac{\varepsilon^4}{16}|C_1| $ such that $|\bar{d}(A^{'},\, B^{'})-\bar{d}(C_s, C_t)| \geq \varepsilon^4$. \label{step2}
\item Count regular pairs: if there are at most $\varepsilon \binom{k_i}{2}$ pairs that are not $\varepsilon$-regular, then stop. $\mathcal{P}_i$ is an $\varepsilon$-regular partition. \label{step3}
\item Refine: apply a refinement algorithm and obtain a partition $\mathcal{P}'$ with $1+k_i4^{k_i}$ classes, such that $sze\_ind(\mathcal{P '}) > sze\_ind(\mathcal{P})$. \label{step4}
\item Go to step 2.
\end{enumerate}

Even if the above mentioned algorithm has polynomial worst case complexity in the size of $G$, there is a hidden tower-type dependence on an accuracy parameter. Unfortunately, Gowers \cite{Gowers1998} proved that this tower function is necessary in order to guarantee a regular partition for \emph{all} graphs. This implies that, in order to have a faithful approximation, the original graph size should be astronomically big. This has typically discouraged researchers from applying regular partitions to practical problems, thereby confining them to the purely theoretical realm.

To make the algorithm truly applicable, \cite{Sperotto07}, and later \cite{Sar+12,Fiorucci17}, instead of insisting on provably regular partitions, proposed a few simple heuristics that try to construct an approximately regular partition. In the next section, we present a new heuristic algorithm which is characterized by an improvement of the summary quality both in terms of reconstruction error and of noise filtering.  
\section{The Summarization Algorithm}\label{sumAlg}
The main limitations which prevent the application of Alon et al.'s algorithm to practical problems concern Step \ref{step2} and Step \ref{step4}. In particular, in Step \ref{step2} the algorithm checks the regularity of all classes pairs by using the three conditions previously described. Given a pair $(C_r,C_s)$, condition \ref{cond1} verifies if it is $\varepsilon$-regular, otherwise conditions \ref{cond2} and \ref{cond3} are used to obtain the \emph{certificates} $C_r^{'}$ and $C_s^{'}$ that witness the irregularity. The main obstacle concerning the implementation of condition \ref{cond3} is the necessity to scan over \textit{almost all possible subsets} of $C_s$. To make the implementation of condition \ref{cond3} feasible, given a class $C_s$, we select in a \textit{greedy way} \label{greedy} a set $Y^{'} \subseteq C_s$ with the highest deviation $\sigma(Y^{'})$ (the deviation is defined in \ref{sigma}). To do so, the nodes of $C_s$ are sorted by bipartite degree, and $Y^{'}$ is built by adding $\frac{\varepsilon^4}{4}n$ nodes with the highest degree. At each iteration of the greedy algorithm, the node with a degree that deviates most from the average degree is added to the candidate certificate $Y^{'}$. This last operation is repeated until the subset $C_s'$, that satisfies condition \ref{cond3}, is found. This almost guarantees to put in the candidate certificate the nodes that have a \textit{connectivity pattern} that deviates most from the one characterizing the majority of the nodes which belong to $C_s$.

As far Step \ref{step4} is concerned, here an irregular partition $\mathcal{P}^{i}_\varepsilon$ is \textit{refined} by a new partition $\mathcal{P}^{i+1}_\varepsilon$, such that the partition index \textit{sze\_idx} is increased. This step poses the main obstacle towards a practical version of Alon et al.'s algorithm involving the creation of an exponentially large number of subclasses at each iteration. Indeed, as we have said, Step \ref{step2} finds all possible irregular pairs in the graph. As a consequence, each class may be involved with up to $(k_i-1)$ irregular pairs, $k_i$ being the number of classes in the current partition $\mathcal{P}^{i}_\varepsilon$, thereby leading to an exponential growth.
To avoid the problem, for each class, one can limit the number of irregular pairs containing it to at most one, possibly chosen randomly among all irregular pairs. This simple modification allows one to divide the classes into a constant, rather than exponential, number of subclasses $l$ (typically $ 2 \leq l \leq 7$).
Despite the crude approximation this seems to work well in practice.         

The devised algorithm takes as input two main parameters, the tolerant parameter $\varepsilon$ and the minimum compression rate $c\_min$, that acts as a stopping criterion in the refinement process, and returns an approximated $\epsilon$-regular partition. Its pseudocode is reported in Algorithm \ref{algorithm:alg1}, where the procedure \textsc{ApproxAlonCertificates}, based on the two heuristics described above, takes as input a partition  $\mathcal{P}_\epsilon^{i}$ and returns the number of irregular pairs of $\mathcal{P}_\epsilon^{i}$. In the next paragraph, we describe the \textsc{Refinement} procedure which refines a partition  $\mathcal{P}_\epsilon^{i}$ into a partition $\mathcal{P}_\epsilon^{i+1}$. Its pseudocode is reported in Algorithm \ref{algorithm:newrefinement}.  
The overall complexity of our summarization algorithm is $O(M(n)) = O(n^{2.376})$, which is dominated by the verification of condition \ref{cond3}.
\begin{algorithm} 
	\caption{The Summarization Algorithm}
	\begin{algorithmic}[1] 
    	\Procedure{ApproxAlon}{$\varepsilon$, c\_min, $G=(V,E)$}
        	\State partitions = empty list
				\State $\mathcal{P}^1_\varepsilon =$ Create initial random partition from $G$
                 		
          \While {True}
          	  \State \#irr\_pairs = $\Call{ApproxAlonCertificates}{\mathcal{P}^i_\varepsilon}$
              \If{\#irr\_pairs $> \varepsilon \binom{k}{2}$ or $\Call{CompressRate}{\mathcal{P}^i_\varepsilon} <$ c\_min}

                	\State break
              \Else
              		\State $\mathcal{P}^{i+1}_\varepsilon = \Call{Refinement}{\mathcal{P}^i_\varepsilon}$
                    \If{$\mathcal{P}^{i+1}_\varepsilon$ is $\varepsilon$-regular}
                    	\State partitions.add($\mathcal{P}^{i+1}_\varepsilon$)
                    \Else
                    	\State break
                    \EndIf
              \EndIf
              \EndWhile
   
            \State Select best partition $\mathcal{P}^*$ with maximum $sze\_idx$ from list partitions
		\EndProcedure
	\end{algorithmic} \label{algorithm:alg1}
\end{algorithm}

\begin{algorithm} 
	\caption{Refinement step performed at the $i$-th iteration of the summarization algorithm 1. Statements 5,10 and 12 may add a node to $C_0$. $\mathcal{P}^{i}_\varepsilon$ is the partition at iteration $i$ of the summarization algorithm}
	\begin{algorithmic}[1] 
		\Procedure{refinement}{$\mathcal{P}^{i}_\varepsilon$}
			\For {each class $C_i$ in $\mathcal{P}^{i}_\varepsilon$}
				\If{$C_i$ is $\varepsilon$-regular with all the other classes}
					\State $C_i =$ \Call{sort\_by\_indegree}{$C_i$}
					\State $C^1_i,C^2_i = $\Call{unzip}{$C_i$}
                \Else
					\State Select $C_j$ with most similar internal structure 
					\State Get certificates $(A^{'}, B^{'})$ and complements $(\bar{A^{'}},\bar{B^{'}})$ of $C_i, C_j$
					\If{$d(A^{'},A^{'})<0.5$}
                    	\State $C^1_i,C^2_i =$ \Call{sparsification}{$A^{'},\bar{A^{'}} \cup \bar{B^{'}}$}
                        
                    \Else
                    	\State $C^1_i,C^2_i =$ \Call{densification}{$B^{'},\bar{B^{'}} \cup \bar{B^{'}}$}
                        
                    \EndIf
					\State Perform step 9,10,11,12 for $B^{'}$
				\EndIf
			\EndFor
			\If{$|C_0| > \varepsilon n$ and $|C_0|>|\mathcal{P}^{i+1}_\varepsilon|$}
            	\State Uniformly distribute nodes of $C_0$ between all the classes
            \Else
            	\State \Return $(\mathcal{P}^{i+1}_\varepsilon, irregular)$
            \EndIf
            \State \Return $(\mathcal{P}^{i+1}_\varepsilon, regular)$
		\EndProcedure
	\end{algorithmic} \label{algorithm:newrefinement}
\end{algorithm}

\indent Given a partition $\mathcal{P}_\epsilon^i = \{C_0,C_1,\dots, C_{k_i}\}$, the \textsc{Refinement} procedure starts by randomly selecting a class $C_i$, then iteratively processes all the others. 
\begin{itemize}
	\item If $C_i$ is $\varepsilon$-regular with all the others, the procedure sorts the nodes of $C_i$ by their \emph{internal degree}, i.e. the degree calculated with respect to the nodes of the same class, obtaining the following sorted sequence of nodes $v_1,v_2,v_3,v_4,v_5,v_6, \cdots, v_{|C_i|}$. The next step splits (\textsc{Unzip}) this sequence into two sets $C_i^1 = \{v_1, v_3, v_5, \cdots , n_{|C_i|-1}\}$ and $C_i^2=\{v_2, v_4, v_6, \cdots , v_{|C_i|}\}$. The latter sets are part of the refined partition $\mathcal{P}^{i+1}_\varepsilon$.
	\item If $C_i$ forms an irregular pair with other classes, the heuristic selects the candidate $C_j$ that shares the most similar internal structure with $C_i$ by maximizing $S = d(C_i, C_j) + (1 - |d(C_i, C_i) - d(C_j, C_j)|)$, where $d(C_i,C_i)=e(C_i,C_i)/|C_i|^2$ is the \emph{internal density}.
	
	After selecting the best matching class $C_j$, we are ready to split the pair $(C_i, C_j)$ in 4 new classes $C_i^1, C_i^2,C_j^1,C_j^2$ based on the internal densities of the certificates $C_i'$ and $C_j'$.
	\begin{itemize}
		\item In particular, a \textsc{Sparsification} procedure is applied when the internal density of a certificate is below a given threshold. This procedure randomly splits the certificate into two new classes. In order to match the equi-cardinality property, the new classes are filled up to $|C_i|/2$ by adding the remaining nodes from the corresponding complement. We choose the nodes that share the minimum number of connections with the new classes.
		\item On the other hand, if the internal density of a certificate is above a given threshold, then a \textsc{Densification} procedure is applied. In particular, the heuristics sorts the nodes of the certificate by their internal degree and \textsc{Unzip} the set into two new classes. Also in this case, we fill the new sets up to $|C_{i}|/2$ by adding the remaining nodes from the corresponding complement by choosing the nodes which share the major number of connections with the new classes. 
	\end{itemize}   
\end{itemize}

\section{Graph Search Using Summaries}\label{graphSearch}
In this section, we discuss how to use our summarization framework to efficiently address the graph search problem defined under a similarity measure. The aim of graph search is to retrieve from a database the top-$k$ graphs that are most similar to a query graph. 
\paragraph{\textbf{Problem Definition}}
We consider a graph database $\mathcal{D}$ containing a high number of simple undirected graphs $g_j \in \mathcal{D}, \; j = 1 \dots |\mathcal{D}|$, and, for the sake of generality, we allow the edges to be weighted. 
\begin{problem}[Graph search]
Given a graph database $\mathcal{D} = \{g_1,g_2,\cdots,g_{|D|}\}$, a query graph $q$, and a positive integer $k$, the graph similarity search problem is to find the top-$k$ graphs in $\mathcal{D}$ that are most similar to $q$ according to a similarity measure.
\end{problem}
As far as the similarity measure is concerned, the most used one is the \emph{graph edit distance} ($GED$) due to its generality, broad applicability and noise robustness \cite{Liang17,Zhang10}. However, since the $GED$ computation is NP-hard, it is not suited to deal with large graphs. To overcome this limitation, we use the \emph{spectral distance} \cite{Jin18}, which is computed by comparing the eigenvalues of the two graphs being matched. The choice of this measure is motivated by the work of Van Dam and Haemers \cite{VANDAM03}, who show that graphs with similar spectral properties generally share similar structural patterns. In this paper, we introduce a slightly modified version of the spectral distance to increase its range of applicability to pairs of graphs that violate the assumption of the Theorem 1 in \cite{Jin18}, which assumes a precise order between the eigenvalues of the two graphs being matched. To this aim, we simply compute the absolute value of the difference between the $i$-$th$ eigenvalue of the first graph with the $i$-$th$ one of the second graph.  

\begin{definition}[Spectral distance]
Given two simple undirected weighted\\graphs $G_1 = (V_1,W_1)$ with $|V_1| = n_1$, and $G_2=(V_2,W_2)$ with $|V_2|=n_2$. Let us denote the corresponding spectra as $0=\lambda_1^{(1)} \leq \lambda_2^{(1)}, \leq \cdots \leq \lambda_{n_1}^{(1)} $ and $0=\lambda_1^{(2)} \leq \lambda_2^{(2)}, \leq \cdots \leq\lambda_{n_2}^{(2)}$. We may assume without loss of generality that $n_2 > n_1$. The spectral distance is then defined as follows
\begin{equation}
SD(G_1,G_2,l) = \frac{1}{n_1} \left ( \sum_{i=1}^{l}|\lambda_i^{(2)} - \lambda_i^{(1)}|  + \sum_{i=l+1}^{n1}|\lambda_{i+n_2-k}^{(2)} - \lambda_i^{(1)}| \right )
\end{equation}
\end{definition}
where, $l$ controls which part of the spectra are being matched. In particular, eigenvalues of $G_1$ are compared with the head and tail eigenvalues of the $G_2$.

\paragraph{\textbf{Using The Summaries}}
In our approach, all the graphs contained in a database are summarized off-line, while the query graph is summarized on-line by means of our summarization framework. Thus, graph search can be performed on graph summaries, and this allows us to speed up the search process and to reduce the storage space. In particular, for each graph $g_j$ of a database $\mathcal{D}$, we store two different quantities: the summary $r_j$ of $g_j$ and the eigenvalues $eig_{r_j}$ of $r_j$.
We then summarized on-line the query graph $q$ obtaining its summary $r_q$. Finally, we compute the spectral distance between $r_q$ and each summary $r_j \in \mathcal{D}$. The desired top-$k$ graphs will be obtained by selecting, from $\mathcal{D}$, the $k$ graphs corresponding to the $k$ smallest value of the spectral distance previously computed. The pseudocode of our approach to graph search is reported in Algorithm $3$.     
\begin{algorithm}
    \caption{Graph Search Using The Summaries}
    \begin{algorithmic}[1]
      \Procedure{AddGraphToDatabase}{$g, \mathcal{D}$}
        
            \State $r =$ Summarize $g$
            \State $eig_r =$ Calculate the eigenvalues of the adj. matrix of $r$
            \State Store $(r, eig_r)$ in $\mathcal{D}$
        \EndProcedure

        \Procedure{2-StageGraphSearch}{$q, \mathcal{D}$}
            \State $r_q =$ Summarize $q$
            \State $eig_{r_q} =$ Calculate the eigenvalues of the adj. matrix of $r_q$
        	\State $sd\_array = \emptyset$
            \For{$r_j$ in $\mathcal{D}$}
                \State $sd =$ Spectral Distance$(r_j, r_q, eig_{r_j}, eig_{r_q})$	   			
                \State Append $sd$ to $sd\_array$
            \EndFor
            \State Order $sd\_array$
            \State \Return first $k$ results of $sd\_array$ and their relative graphs.        
        \EndProcedure
    \end{algorithmic} \label{gS}
\end{algorithm}
\section{Experimental Evaluation}\label{experiments}
In this section, we evaluate our summarization algorithm both on synthetic graphs and on real-world networks to assess:
\begin{itemize}
\item the ability of the proposed algorithm to separate structure from noise;
\item the usefulness of the summaries in retrieving from a database the top-$k$ graphs that are most similar to a query graph.
\end{itemize}

\begin{figure}[t!]
  \centering
  \includegraphics[width=0.43\linewidth]{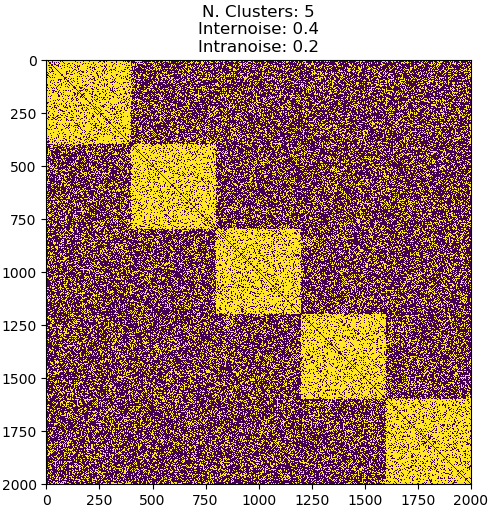}
  \caption{The adjacency matrix of an undirected synthetic graph of 2000 vertices. The graph is generated by corrupting $5$ cliques as described in Algorithm \ref{algorithm:gen}. In particular, the \emph{intra-cluster noise probability} is $0.2$ and the \emph{inter-cluster noise probability} is $0.4$.}
\label{fig:graphex}
\end{figure}

\paragraph{\textbf{Experimental Settings}}
In our experiments we used both synthetic graphs and real-world networks. We generated synthetic graphs with a cluster structure, where the clusters are perturbed with different levels of noise. In particular, each graph is generated by adding spurious edges between cluster pairs and by dropping edges inside each cluster. Figure \ref{fig:graphex} provides a concrete example with a visual explanation. The pseudocode of the algorithm used to generate the synthetic datasets is reported in Algorithm \ref{algorithm:gen}.

As far as the real-world networks are concerned, we used two different datasets which have been taken from two famous repositories: the Stanford Large Network Dataset Collection SNAP \cite{snapnets} and the Konect repository of the University Koblenz-Landau Konect \cite{konect}. In particular, we used the following networks: Facebook \cite{facebookNIPS}, Email-Eu-core \cite{Yin2017LocalClustering}\cite{Leskovec07email}, Openflights\cite{konect:2016:openflights}, and Reactome \cite{Joshi-Tope2005Reactome:Pathways}. Our algorithm is implemented in Python $3.6.3$  \footnote{The implementation is available from \href{ }{https://github.com/MarcoFiorucci/graph-summarization-using-regular-partitions}} and the experiments are performed on an Intel Core i5 @ 2.60GHz HP Pavilion 15 Notebook with 8GB of RAM (DDR3 Synchronous 1600 MHz) running Arch-Linux with kernel version 4.14.4-1.

\begin{algorithm}
    \caption{Synthetic graph generator. \emph{Input parameters}: $n$ is the size of the desired graph $G$; $num\_c$ is the number of clusters contained in $G$; $\eta_1$ is the probability of adding a spurious edge between a pair of clusters (inter-cluster noise probability); $\eta_2$ is the probability of dropping an edge inside a cluster (intra-cluster noise probability). \emph{Output}: $G$.}
    \begin{algorithmic}[1]
        \Procedure{SynthGraphGen}{$n, num\_c,\eta_1, \eta_2$}
        
            \State $G =$ Generate Erd\H{o}s R\'enyi graph of size $n$ using  $\eta_1$ as edge probability 
            \State $clust\_dim = n / num\_c$
            \For{$i$ in $num\_c$}
                \State Select $clust\_dim$ nodes from $G$ and create cluster $c_i$ with them
                \State For each edge in $c_i$ drop it with probability $\eta_2$
            \EndFor
            \State \Return $G$
       
        \EndProcedure
    \end{algorithmic}
    \label{algorithm:gen}
\end{algorithm}

\subsection{Graph Summarization}
We performed experiments on both synthetic graphs and on real-world networks to assess the ability of the proposed algorithm to separate structure from noise. As evaluation criterion, we used the reconstruction error, which is expressed in terms of normalized $l_p$ norm computed between the similarity matrix of an input graph $G$ and the similarity matrix of the corresponding reconstructed graph $G'$.

\begin{definition}[Reconstruction error] 
	Given the similarity matrix of the input graph $A_G$ and the similarity matrix of the reconstructed graph $A_{G'}$, the reconstruction error is defined as follows: 
	\[l_p(\mathbf{A_G}, \mathbf{A_{G'}}) = (\sum_{i=1}^n \sum_{j=1}^n (A_G(i,\,j) - A_{G'}(i,\,j))^p)^\frac{1}{p}\] 
\end{definition}

We decided to use the reconstruction error in order to compare our results with the ones presented by Riondato et al. \cite{Riondato2017}, who evaluated the summary quality using this measure. This choice is due to the fact that their algorithm summarizes a graph by minimizing the reconstruction error. However, they pointed out that the reconstruction error has some shortcomings. In particular, given an unweighted graph $G$, it is possible to produce an uninteresting summary with only one supenode corresponding to the vertex set and $l_1$ reconstruction error at most $n^2$. On the other hand, if we obtained an useful summary, where each pair of vertices belonging to a supernode share an high number of common neighbors, then we get a low (say $o(n^2)$) $l_1$ reconstruction error: this is a desirable behavior because low values of $l_1$ correspond to high quality summaries. Unfortunately, such low values are often obtained only with summaries having an high number of supernodes. This prevents to adopt the reconstruction error as a general measure to assess the summary quality.

\begin{figure}[t!]
	\centering
	\includegraphics[width=1\linewidth]{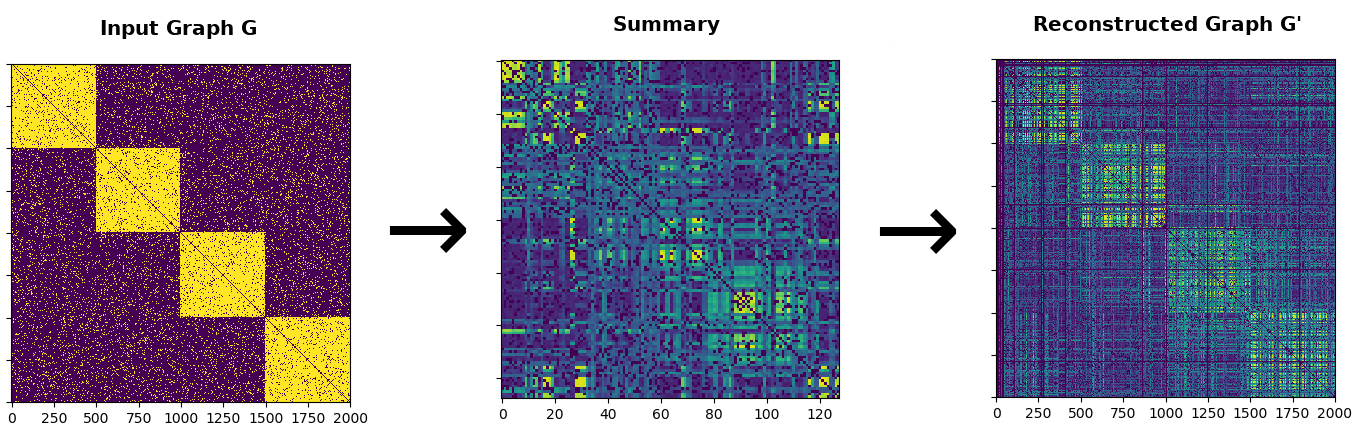}
	\caption{We summarized an input graph $G$ by using Algorithm \ref{algorithm:alg1}. We then "blow up" the summary to obtain the reconstructed graph $G'$.}
	\label{fig:summarization}
\end{figure}

As far the summarization and reconstruction steps are concerned, we proceeded, in all the experiments, in the following way: we applied our summarization algorithm (see Algorithm \ref{algorithm:alg1}) to summarize an input graph $G$. We then ``blow-up'' the summary in order to obtain the reconstructed graph $G'$, which preserves the main structure carried by the input graph (Figure \ref{fig:summarization}).

\paragraph{\textbf{Noise Robustness Evaluation}}  
We study the ability of the proposed algorithm to separate structure from noise in graphs performing an extensive series of experiments on both synthetic graphs and real-world networks. As far synthetic graph experiments are concerned, we generated a graph $G$ by corrupting the clusters of $GT$ in the following way: we added spurious edges between each cluster pair with probability $\eta_1$, and we dropped edges inside each cluster with probability $\eta_2$ (see algorithm \ref{algorithm:gen}). As far as the real-world networks experiments are concerned, we added spurious edges with probability $noise$ $probability$ to a original graph $GT$ obtaining an input graph $G$.

In the ideal case, the distance between $G$ and the corresponding reconstructed graph $G'$ should be only due to the filtered noise, while the distance between $GT$ an $G'$ should be closed to zero. Hence, we computed the reconstruction error $l_2(G',GT)$ to assess the robustness of our summarization framework against noise.  

\begin{figure}[t!]
	\centering
	\includegraphics[width=0.5\linewidth]{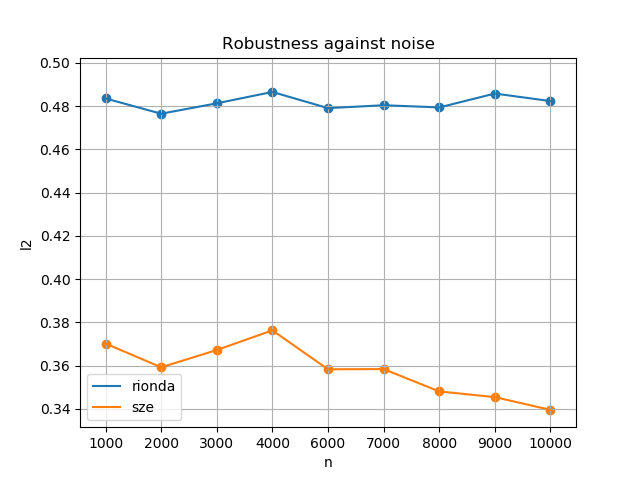}
	\caption{The plot shows the medians computed, for each size $n$, from the $25$ values of $l_2(G'_{n,(\eta_1,\eta_2)},GT)$, where $(\eta_1,\eta_2)$  corresponds to one of the $25$ pairs of noise probabilities. The curve "rionda" is obtained by using Riondato et al.'s algorithm \cite{Riondato2017}, while the curve "sze" is obtained by applying our summarization framework.} 
 	\label{fig:l2vsnoise}
\end{figure}

\emph{Experiment 1.} We generated synthetic graphs of different sizes, spanning from $10^3$ up to $10^4$ nodes. We synthesized $250$ graphs by considering, for each of the $10$ different sizes, all the $25$ combinations of the following noise probabilities:
\begin{itemize}
\item the probability $\eta_1$ of adding a spurious edge between a pair of clusters, called \emph{inter-cluster noise probability}, which assumes values in $\{0.1,\,0.2,\,0.3,$ $\,0.4,\,0.5\}$;
\item the probability $\eta_2$ of dropping an edge inside each cluster, called \emph{intra-cluster noise probability}, which assumes values in $\{0.1,\,0.2,\,0.3,\,0.4,\,0.5\}$.
\end{itemize}
Let's consider a synthetic graph $G_{n,(\eta_1,\eta_2)}$, where n is its size, and $(\eta_1,\eta_2)$ corresponds to one of the $25$ pairs of noise probabilities. For each $G_{n,(\eta_1,\eta_2)}$ we obtained the reconstructed graph $G'_{n,(\eta_1,\eta_2)}$, and we then computed the reconstruction error $l_2(G'_{n,(\eta_1,\eta_2)},GT)$. Given a size $n$, we computed the median $m_n$ of $ \{  l_2(G'_{n,(0.1,0.1)},GT), \;   l_2(G'_{n,(0.1,0.2)},GT), \cdots ,$ \\ $  l2_(G'_{n,(0.5,0.5)},GT) \}$. We reported in figure \ref{fig:l2vsnoise} the $10$ medians computed using our summarization framework and the corresponding medians obtained by applying Riondato et al.'s algorithm \cite{Riondato2017}. We can see how our framework outperforms the state-of-the-art summarization algorithm in terms of robustness against noise.  

\emph{Experiment 2.} The aim of this experiment is to study separately the robustness against the inter-cluster and the intra-cluster noise. Let's consider the probability of dropping an edge inside each cluster $\eta_2$ equals to $0.2$ and the graph size $n$ equals to $10^4$. To asses the inter-cluster noise robustness, we generated synthetic graphs $G_{10^4,(\eta_1,\,0.2)}$, where $\eta_1$ assumes values in $\{0.1,\,0.15,\,0.2,\,0.25,\,0.3,$ $\, 0.35,\,0.4,\,0.45,\,0.5\}$, and we computed the reconstruction errors $l_2(G'_{10^4,(\eta_1,\,0.2)},$ $GT)$. As far the intra-cluster noise is concerned, we chose the probability of adding a spurious edges between each pair of clusters $\eta_1=0.2$, and the graph size $i=10^4$. We then generated synthetic graphs $G_{10^4,(0.2, \, \eta_2)}$, where $\eta_2$ assumes values in $\{0.1,\,0.15,\,0.2,\,0.25,\,0.3,$ $\, 0.35,\,0.4,\,0.45,\,0.5\}$, and we computed the reconstruction errors $l_2(G'_{10^4,(0.2, \, \eta_2)},GT)$.

\begin{figure}[!ht]
	\centering
	\includegraphics[width=0.45\linewidth]{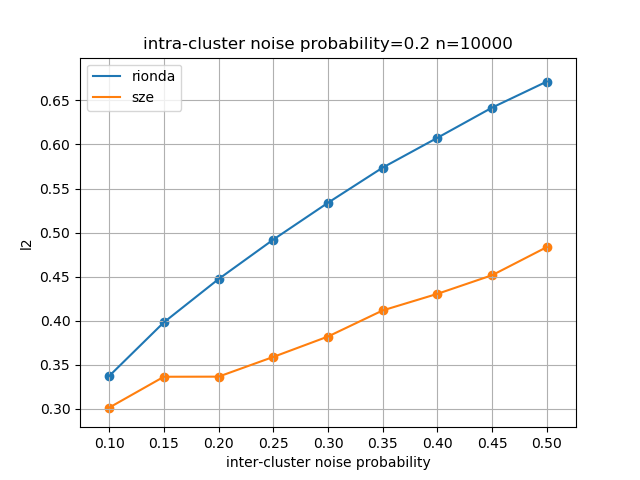}
    \includegraphics[width=0.45\linewidth]{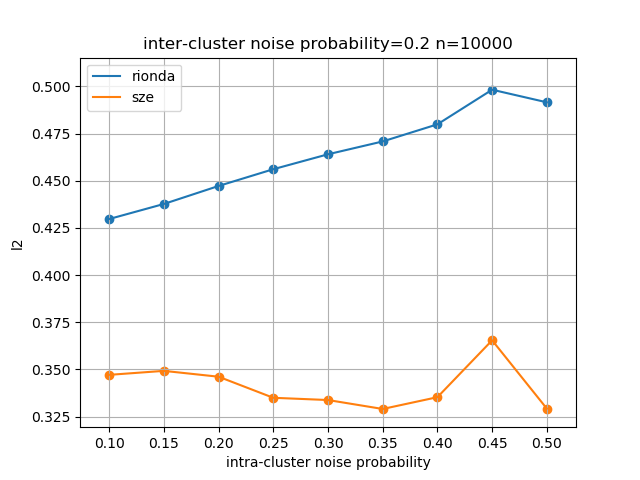}
	\caption{The plot on the left represents $l_2(G'_{10^4,(0.2,\,\eta_2)},GT)$ versus the intra-noise probability. The plot on the right represents $l_2(G'_{10^4,(\eta_1,\,0.2)},GT)$ versus the inter-noise probability. The curve ``rionda'' is obtained by using Riondato et al.'s algorithm, while the curve ``sze'' is obtained by applying our summarization framework.} 
 	\label{fig:l2_vs_inter}
\end{figure}

Figure \ref{fig:l2_vs_inter} illustrates the comparison between our results with those obtained by applying Riondato et al.'s algorithm \cite{Riondato2017}. This results are in accord to those presented in figure \ref{fig:l2vsnoise}, and provides an experimental verification of the ability of our method to separate structure from noise in graphs.    

\begin{figure}[!ht]
	\centering
	\includegraphics[width=0.45\linewidth]{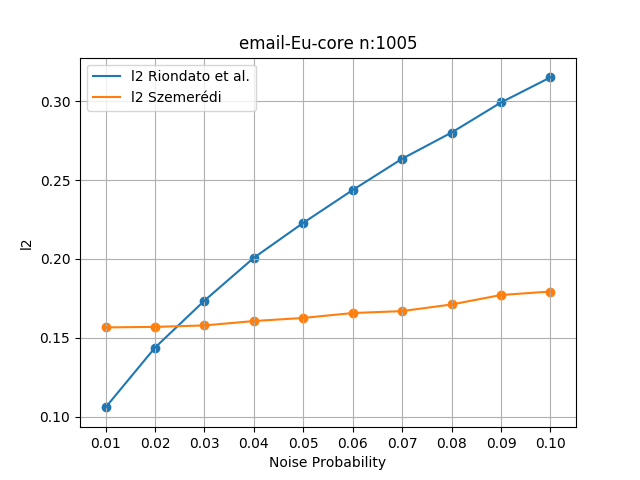}
    \includegraphics[width=0.45\linewidth]{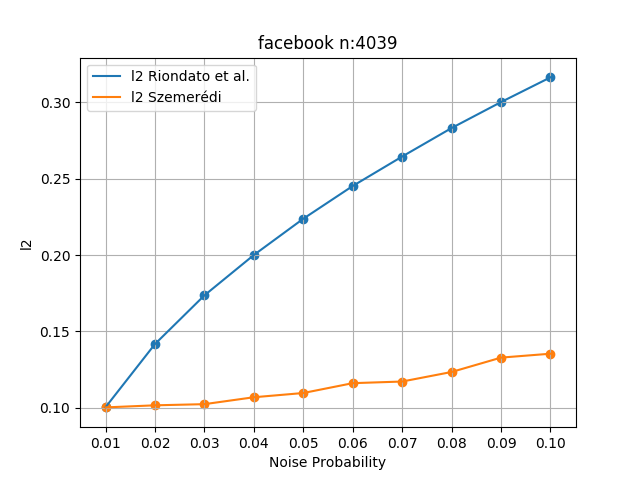}\\
    \includegraphics[width=0.45\linewidth]{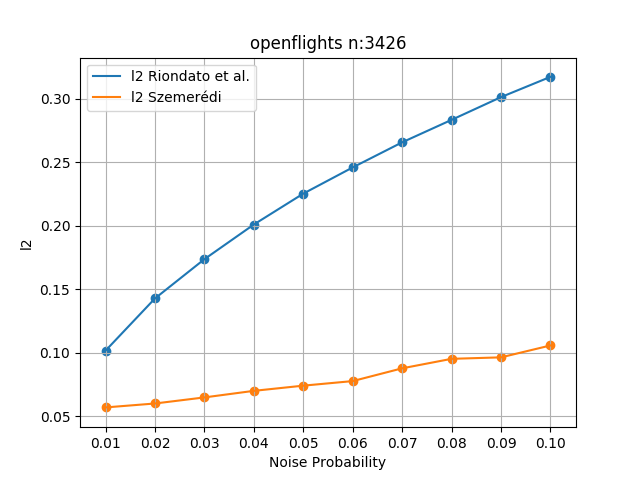}
    \includegraphics[width=0.45\linewidth]{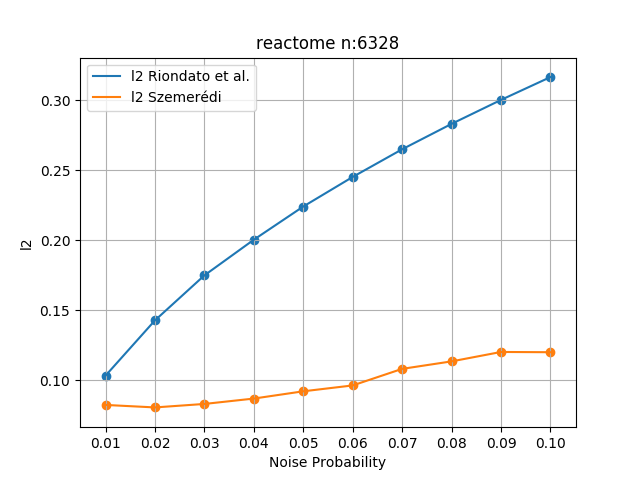}
    \caption{These plots represent the median of the $l_2(G',GT)$ versus the noise probability. We run $20$ experiments for each value of the  noise probability. The curve ``rionda'' is obtained by using Riondato et al.'s algorithm, while the curve ``sze'' is obtained by applying our summarization framework.} 
 	\label{fig:l2_real-wordl}
\end{figure}

\emph{Experiment 3.} We added spurious edges with probability $noise$ $probability$ to an original real-world network $GT$ obtaining an input graph $G$. The $noise$ $probability$ assumes values in $\{0.01, \, 0.02,\,0.03,\,0.04,$ $\, 0.05,\,0.06,\,0.07,\,0.08, \, 0.09, \,$  $0.1\}$. We applied this procedure on real-word networks, which have been taken from the Stanford Large Network Dataset Collection SNAP \cite{snapnets} and from the Konect repository of the University Koblenz-Landau Konect \cite{konect}.
Since our framework is based on the Regularity Lemma, which is suited to deal only with dense graphs, we expect to obtain low quality summaries from sparse real-world networks. However, as shown in figure \ref{fig:l2_real-wordl}, our framework outperforms the state-of-the-art summarization algorithm in terms of robustness against noise providing good quality summary even on sparse real-world networks. In particular, we can see how the quality increases with the size of the input graph, which is in accord with the assumptions of the Regularity Lemma.

\subsection{Graph Search} We performed extensive experiments on synthetic datasets to assess the usefulness of the summaries in retrieving, from a database, the top-$k$ graphs that are most similar to a query graph. To this end, we evaluate the quality of the answer in terms of the found top-$k$ similar graphs, and we evaluate the scalability both in the size of the database and in the size of the graphs.    
\paragraph{\textbf{Quality Evaluation}}\label{exp:quality-eval}
We conducted the following experiment: we compared graph search on the summaries with the baseline approach, in which the spectral distance is computed between no preprocessed graphs. The aim of the experiment is to show that pre-summarizing the graphs in the databases increases the noise robustness of the search process.
We created a database $\mathcal{D}$ contained synthetic graphs, which have different structures corrupting with different levels of noise (see algorithm \ref{algorithm:gen}). In particular, each graph is generated by combining the following three factors: five different possible number of clusters $\{4, 8, 12, 16, 20\}$, six different possible levels of intra-cluster noise probability $\{0.05, 0.1, 0.15, $ $0.2, 0.25, 0.3\}$ and six different possible levels of inter-cluster noise probability $\{0.05, 0.1, 0.15, 0.2, 0.25,$ $ 0.3\}$. Given a size $n$, we generated $180$ graphs considering all the possible combinations of these three parameters. As described in Algorithm \ref{gS}, we stored in $\mathcal{D}$ the eigenvalues of the $180$ synthetic graphs, their summaries and the corresponding eigenvalues. Finally, we grouped the database graphs into five groups. Each group $\omega_i$, with $i=1,2,3,4,5$, is composed by 36 graphs that are generated by corrupting the same cluster structure with different combinations of intra-cluster and inter-cluster noise probability. Hence, all the graphs belonging to a given group $\omega_i$ are similar, since they have the same main structure. 

More formally, we constructed a set $Q$ of five query graphs by randomly sampling one graph from each group $\omega_i$. Then, we first computed the spectral distance between $q_i \in Q$ and every graph in the database $\mathcal{D}$. We then calculated the $AP@k$ for each query $q_i$ by considering relevant the graphs belonging to $\omega_i$ i.e. the same group of $q_i$. Finally, we computed the $MAP@k$ score by averaging the \emph{average precision} $AP@k(q_i)$ of the five graphs in $Q$. 

We repeated the same procedure using the summaries of the $180$ synthetic graphs contained in $\mathcal{D}$. The aim of this experiment is to compare the quality obtained using our approach with that obtained by computing the spectral distance between original graphs. 
We performed the experiment by considering the following graph sizes $n = 1500, 2000, 3000, 7000$.\\ 

\noindent The $AP@k(q_i)$ and the $MAP@k$ are defined as follows.

\begin{definition}[Average precision]
Given a query $q \in Q$, a set of relevant graphs $\omega_i$ (graphs that share the same structure with $q$). Let us consider the output top-$k$ graphs in a database $\mathcal{D}$ ordered by crescent spectral distance. We define the average precision at $k$ as follows.
\begin{equation}
AP@k(q) = \frac{1}{|w_i|} \cdot \sum^{k}_{j=1} Precision(j) \cdot Relevance(j)
\end{equation}
where $Precision(j)$ is the relevant proportion of the found top-$k$ graphs, while $Relevance(j)$ is $1$ if the considered graph is part of $\omega_i$ and is $0$ otherwise. Finally, $|\omega_i|$ is the number of relevant graphs. 
\end{definition}

\begin{definition}[Mean average precision]
Given a query set $Q$, the mean average precision is defined as follows.
\begin{equation}
MAP(Q) = \frac{1}{|Q|} \cdot \sum_{q_i \in Q} AP@k(q_i)
\end{equation}
\end{definition}

\begin{figure}[t!]
  \centering
  \includegraphics[width=0.5\linewidth]{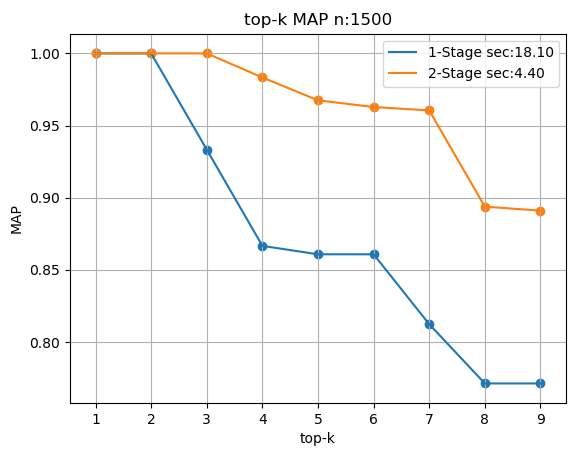}
  \caption{The MAP@k of the top-$k$ graphs given as output in a database of 180 graphs. The size of the graphs contained in the database is $n = 1500$. Two-stage is referred to the use of our pre-summarization approach to address the graph search problem, while one-stage is referred to the search on the original graphs.}
\label{fig:ranking1}
\end{figure}

\begin{figure}[!ht]
  \centering
  \includegraphics[width=0.5\linewidth]{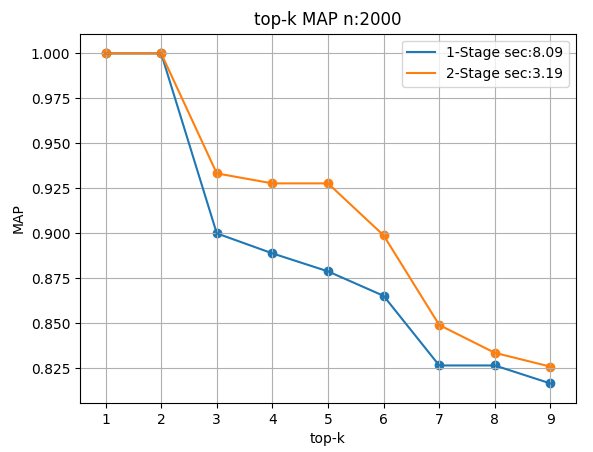}
  \caption{The MAP@k of the top-$k$ graphs given as output in a database of 180 graphs. The size of the graphs contained in the database is $n = 2000$. Two-stage is referred to the use of our pre-summarization approach to address the graph search problem, while one-stage is referred to the search on the original graphs.}
\label{fig:ranking2}
\end{figure}

\begin{figure}[!h]
  \centering
  \includegraphics[width=0.5\linewidth]{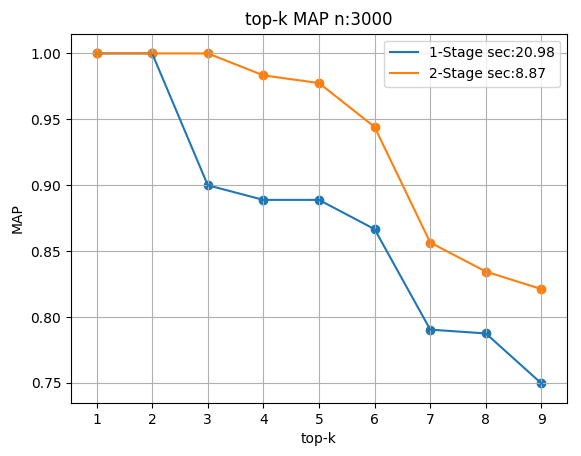}
  \caption{The MAP@k of the top-$k$ graphs given as output in a database of 180 graphs. The size of the graphs contained in the database is $n = 3000$. Two-stage is referred to the use of our pre-summarization approach to address the graph search problem, while one-stage is referred to the search on the original graphs.}
\label{fig:ranking3}
\end{figure}

\begin{figure}[!ht]
  \centering
  \includegraphics[width=0.5\linewidth]{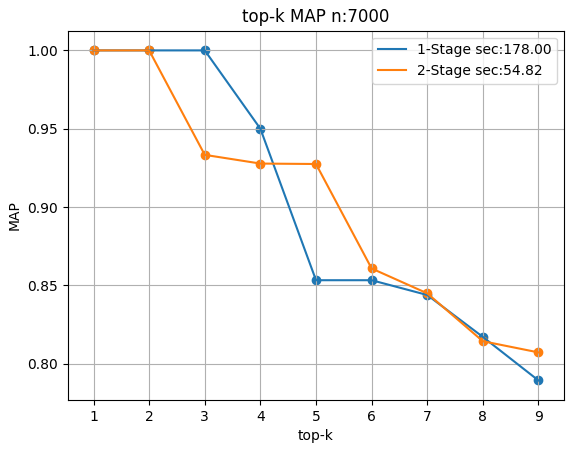}
  \caption{The MAP@k of the top-$k$ graphs given as output in a database of 180 graphs. The size of the graphs contained in the database is $n = 7000$. Two-stage is referred to the use of our pre-summarization approach to address the graph search problem, while one-stage is referred to the search on the original graphs.}
\label{fig:ranking4}
\end{figure}

In particular, the higher is the value of the $MAP \in [0,1] $, the higher is the quality of the proposed graph search algorithm. Figures \ref{fig:ranking1}, \ref{fig:ranking2}, \ref{fig:ranking3}, \ref{fig:ranking4} show that the proposed summarization based approach improved the query quality.      

\paragraph{\textbf{Scalability}}\label{exp:scalability}
In order to evaluate the scalability of our approach, we conducted two different experiments. In the first one, we investigated the time required to perform a single query as the dimension of the database $\mathcal{D}$ grows. In the second one, we investigated the query time in function of the size of the query graph.

In the first experiment, we fixed the size of all the graphs to be $n=2000$. We then generated the graphs in $\mathcal{D}$ using all the possible combinations of the following factors: three different numbers of clusters $\{4, 12, 20\}$, six different levels of intra-cluster noise probability $\{0.05, 0.1, 0.15, $ $ 0.2, 0.25, 0.3\}$, and six different levels of inter-cluster noise probability $\{0.05, 0.1, 0.15, 0.2, 0.25,$ $ 0.3\}$. The combination of these three parameters allow us to generate 108 graphs. We then copied them enough times to reach a database cardinality spanning from $10^3$ up to $10^4$ graphs.

\begin{figure}[t!]
  \centering
  \includegraphics[width=0.5\linewidth]{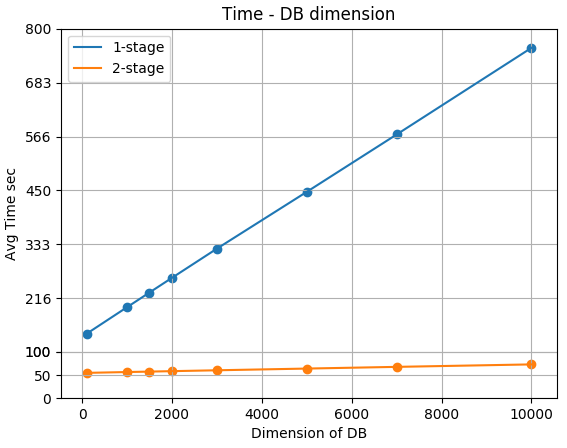}
  \caption{This plot shows the time expressed in seconds to perform a query as we increase the size of the database. Two-stage is referred to the use of our pre-summarization approach to address the graph search problem, while one-stage is referred to the search on the original graphs.}
  \label{fig:scalability1}
\end{figure}

The query time is calculated as follows: 
\begin{equation}
t = t\_s(q) + t\_{eig(r_q)} + t\_SD(eig_{r_q}, eig_{r_j})  \;\;\;\;\; j=1,\dots, |\mathcal{D}|.
\end{equation}

where $t\_s(q)$ is the time required to obtain the summary $r_q$  of the query graph $q$;  $t\_{eig}(r_q)$ is the time required to calculate the eigenvectors of $r_q$; and  $t\_SD(eig_{r_q}, eig_{r_j})$ is the time required to calculate the spectral distances between $r_q$ and each graph summary $r_j$ contained in $\mathcal{D}$.
We reported in Figure \ref{fig:scalability1}, the computed time $t$ versus the cardinality of the database $\mathcal{D}$.

\begin{figure}[t!]
  \centering
  \includegraphics[width=0.5\linewidth]{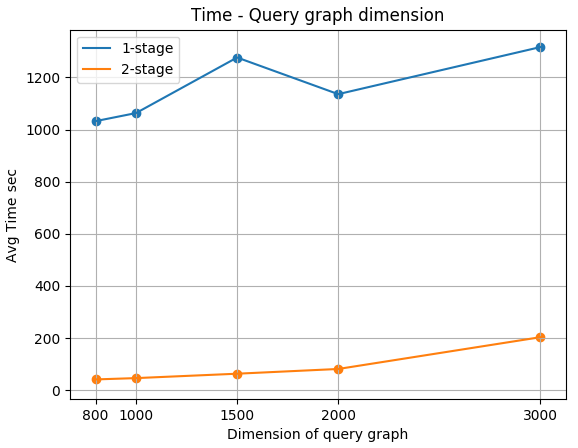}
  \caption{This plot shows the time (expressed in minutes) for retrieving the top-$k$ graphs in a database composed of 10000 graphs as we increase the dimension of the query graph. Two-stage is referred to the use of our pre-summarization approach to address the graph search problem, while one-stage is referred to the search on the original graphs.}
  \label{fig:scalability2}
\end{figure}

In the second experiment, we generated different databases $\mathcal{D}_i$, containing $10000$ graphs. All the graphs in $\mathcal{D}_i$ have the same size and have been created analogously as the previous experiment. We then constructed a query graph $q_i$ of the same size of the graphs in $\mathcal{D}_i$, and we measured the query time $t_i$ as we did for the previous experiment. We reported in Figure \ref{fig:scalability2}, the computed $t_i$ versus the size of the graph query $q_i$.
Figures \ref{fig:scalability1}, \ref{fig:scalability2} provide us an experimental verification of the scalability of our approach both in the size of the database and in the size of the query graph.

\section{Conclusions}\label{conclusions}
In this work, we introduced a graph summarization approach based on the Regularity Lemma, which provides us with a principled way to describe the essential structure of large graphs using a small amount of data. We have successfully validated our framework both on synthetic and real-world graphs showing that our algorithm surpasses the state-of-the-art graph summarization methods in terms of noise robustness. 
In the second part of the paper, we presented an algorithm to address the graph similarity search problem exploiting our summaries. In particular, the proposed method is tailored for efficiently dealing with databases containing a high number of large graphs, and, moreover, it is robust against noise, which is always presented in real-world data. This achievement seems of particular interest since, to the best of our knowledge, we are the first to devise a graph search algorithm which satisfies all the above requirements together. 
A weak point of our summarization algorithm is related to its time complexity, which prevents the application of our framework to networks of millions of nodes. This demands for the development of efficient approaches suited to deal with large sparse graphs. In this direction, it would be a good idea to develop a heuristic based on the version of the Weak Regularity Lemma introduced by Fox et al. \cite{Fox2018}, as well as designing a distributed version of the regular decomposition algorithm introduced by Reittu et al. \cite{Reittu2018}, who studied the linkage among the Regularity Lemma, the Stochastic Block Model and the Minimum Description Length. We think that the notion of regular partition will allow us to tackle the scalability issue faced by graph-based approaches \cite{Bunke2011, Vento2015}. This  would  pave  the  way for a principled approach to massive network data analysis by combining modern graph theory and combinatorics with machine learning and pattern recognition.

\section*{References}

\bibliographystyle{splncs04} 
\bibliography{references}


\end{document}